\documentclass[]{RAA}            

\usepackage{graphicx,times}             
\usepackage{natbib}
\usepackage{amssymb,amsmath}
\usepackage{verbatim}
\usepackage{ulem}
\bibpunct{(}{)}{;}{a}{}{,}

\usepackage[pagebackref=true]{hyperref}

\begin{document}

  \title{
  To detect strong nugget with an acoustic array
}

   \volnopage{Vol.0 (20xx) No.0, 000--000}      
   \setcounter{page}{1}          

   \author{Hao-Yang Qi 
      \inst{1,2}
   \and Ren-Xin Xu 
      \inst{1,2}
   }

   \institute{State Key Laboratory of Nuclear Physics and Technology, Peking University, Beijing 100871, China; {\it qhy021@pku.edu.cn,~r.x.xu@pku.edu.cn}\\
        \and
             Department of Astronomy, School of Physics, Peking University,
             Beijing 100871, China\\
\vs\no
   {\small Received 20xx month day; accepted 20xx month day}}

\abstract{
This article discusses strong nuggets (SNs) which means strong interaction condensed matter with a mass of about $10^6\,$g. They may originate from the early universe, supernova, pulsar merger event, and so on. Depending on the equation of state, the SNs could be stable and even be one of the candidates for dark matter. In order to detect SNs which hitting the Earth or the Moon at a non-relativistic velocity, a new messenger, the acoustic array, is analysed. The results of the calculations show that the impact signal of an SN can be detected at a distance of about 30 kilometers from the nugget's trajectory. By using microphone boxes, hydrophones or seismographs to construct an array in the bedrock, ocean or on the Moon, it is possible to reconstruct the velocity, mass, and interacting cross section of SNs, and then constrain also the nature of supra-nuclear matter. The acoustic array can also be used for distributed acoustic sensing of meteorites or earthquakes. The sonar localisation system on the proposed High-energy Underwater Neutrino Telescope (HUNT) is suggested as a pathfinder for acoustic array detection.
\keywords{dark matter, dense matter, shock waves, instrumentation: detectors}}

\authorrunning{H.-Y. Qi \& R.-X. Xu}            
\titlerunning{
   To detect strong nugget with an acoustic array
   }

\maketitle

\section{Introduction}           
\label{sect:intro}

The knowledge humans have about the material world was surrounded by speculation or superstition until the early 20th century when the era of quantum and relativity arrived, and consequently, leading eventually to the great success of the ``standard model of particle physics'' in the 1960s and 1970s, in which the building blocks of the world are fundamental fermions with interactions mediated by gauge bosons.
In contrast to the inductive approach mentioned above, the deductive derivation of the standard model is embraced in contemporary physics, particularly in the study of multiquark states~\citep[e.g.,][]{2016PhR...639....1C,2018RvMP...90a5004G,2018ARNPS..68...17K,2023RPPh...86b6201C}.
This field serves as a critical testing ground for the underlying theory of the strong force, the quantum chromodynamics (QCD), in its non-perturbative regime.
It is well known that normal baryonic matter consists only of atomic nuclei and electrons, condensed by the electromagnetic force in solid or liquid states, while the nuclei are droplets of stable nucleons condensed by the strong interaction.\footnote{%
By convention, we simply call the former electric matter and the latter strong matter. A strong nugget (SN) is actually QCD-condensed matter with a large baryon number $A$ (e.g., $A>10^3$) and at about nuclear density.
}
Could strong nuggets (SNs) made up of other stable bound states (e.g. ``exotic'' baryons) and even free quarks exist in our world? How can we even detect them?
These questions are the focus of this article.

The above questions become more pressing as we try to understand a star when its nuclear power is exhausted.
In the glorious era of the development of quantum mechanics, the new Fermi-Dirac statistics was first applied to understand white dwarfs, with the quantum degenerate pressure against self-gravity~\citep{1926MNRAS..87..114F}, but the pressure cannot be sufficient when the energy-momentum relation is corrected by special relativity, so that the dwarfs must collapse if their masses are greater than a certain limit~\citep{Chandrasekhar1931_ApJ74-8}.
What happens beyond this Chandrasekhar limit?
A neutral ``giant nucleus'' was superficially anticipated before the discovery of the neutron~\citep{Landau1932pZS1-285,2013PhyU...56..289Y}, which can be regarded as the prototype of the neutron star, and the unexpected pulsars~\citep{1968Natur.217..709H} were soon thought to be the physical reality of the speculated neutron objects.
However, in the era of the standard model of particle physics, nucleons (i.e., protons and neutrons) are bound states of quarks,  mostly with three light flavours, and one can in principle have an another natural way to neutralize a gigantic nucleus, besides neutronization, if strangeness is included~\citep[for recent introductions, e.g.,][]{2023AdPhX...837433,Xu2024,Xia-Gao-Xu2024}.
What is then the real nature of pulsar-like compact objects?\footnote{%
These objects are commonly referred to {\em neutron} stars in the literature, but this does not mean that they are actually made up of neutrons. Never take that word for granted.
} %
It is worth noting that, due to the non-perturbative QCD behaviour, the equation of state of the supra-nuclear matter inside pulsars is the most challenging problem in multi-messenger astronomy, especially after the discovery of gravitational waves from binary neutron star mergers~\citep{2019PrPNP.10903714B}.
Nevertheless, roughly neutral SNs would be a direct consequence of the fact that pulsars are strange stars with the light flavour symmetry of valance quarks, and it is highly encouraged to explore the observational implications of these SNs.

There is, of course, a variety of SNs proposed in either the ``old'' physics (i.e., the standard model) or the  new physics (i.e., beyond the standard model).
Collapsed nuclei with strangeness have been conjectured in the quark model~\citep{1971PhRvD...4.1601B}, and strange quark matter (SQM) in bulk is seriously discussed as a ground state of baryonic matter~\citep[e.g.,][]{Witten+1984}, manifesting in the form of compact stars, cosmic rays, and even dark matter.
An axion domain wall was invented to enclose the SQM, which could generate pressure to stabilise it against decay into baryons~\citep{Zhitnitsky+2003,Zhitnitsky+2021}.
These axion-quark nuggets (AQN) could consist of both matter and antimatter, preserving the quark-antiquark symmetry during the QCD epoch. 
There is also an alternative way to keep quark-antiquark symmetry, in which all the antiquarks are simply bound into antiquark nuggets (antiQN)~\citep{Flambaum+etal+2023,Flambaum+etal+2025}.
Nevertheless, non-strange two-flavour quark matter (i.e. $ud$QM) has also been conjectured with the inclusion of flavour feedback on the QCD vaccum, which could be more stable than SQM and ordinary nuclear matter at sufficiently large baryon numbers~\citep{2018PhRvL.120v2001H,2020PhRvD.101d3003Z,2020PhRvD.102h3003R}.
It is worth noting that such a diverse form of SNs is also proposed as a candidate for dark matter in the extremely non-relativistic regime, the enigmatic matter with speculated particle masses of more than ninety orders of magnitude~\citep{2024arXiv240601705C}, i.e., from $10^{-21}$\,eV to $10^{37}$\,g, or ($10^{-30}\sim 10^{61}$)\,GeV.

The story of strong-interaction matter changes, however, if non-perturbative QCD effects for bulk strong matter are treated with care, potentially leading to two consequences: (1) special bound states as a result of ``condensation'' in position space for quark matter, and (2) the enigmatic hard core between the states, as in an analogy to that between nucleons.
Certainly, atomic nuclei are the epitome of the case of two flavours of valance quarks, in which nucleons are the stable bound states.
An extension from two to three flavours would be welcome and straightforward, and the corresponding nucleon-like bound states have been conjectured to be {\em strangeons},\footnote{%
It is evident that strangeons not only form QCD-bound states but also exhibit a short-range repulsion between each other.
Even for atomic nuclei, the nuclear repulsive core, which provides the nuclei the stability, is still not well understood in QCD.
Nonetheless, one might be able to model the hard-core with a linked bag approach~\citep{Miao+2022}.
} %
formerly known as strange clusters~\citep{Xu+2003}.
In addition to the micro-physical logic mentioned above, the idea of strangeon matter (SnM) in bulk would also support Landau's belief of neutralizing a gigantic nucleus.
Indeed, the strangeon star model shows indomitable vitality in understanding astronomical observations, including the stiff equation state of massive pulsars~\citep{LX2009}, the low tidal deformability of binary compact star mergers~\citep{2019EPJA...55...60L}, and the huge free-energy required for burst events~\citep{2006MNRAS.373L..85X,2024RAA....24b5005C,2024RAA....24j5012W}.
In addition, it is proposed that a crossover QCD phase transition could occur in the early universe, and that stable strangeon nuggets (SnNs) could form before the Big Bang nucleosynthesis, manifesting in the form of dark matter~\citep{Wu+2024}.
Anyway, we can sum up different versions of speculated strong matter: SQM, AQN, antiQN, $ud$QM, and SnM, and are focusing so-called ``strong nuggets'' with baryon numbers much smaller than those of stars and planets in this paper.

To be honest, we cannot yet determine the minimum baryon number, $A_{\rm crit}$, a critical value for stable SnNs.
Both the fundamental strong and weak interactions should be included to define this value.
Given the importance of the electron's role in the asymmetry of $e^\pm$~\citep{2020SCPMA..6319531X}, we could therefore use the electron Compton wavelength, $\lambda_{\rm C} = h/(m_{\rm e}c)$, to estimate the critical baryon number, $A_{\rm crit} \sim (\lambda_{\rm C}/1\,{\rm fm})^3 \simeq 10^{10}$.
The value of $A_{\rm crit}$ could become smaller if the strong interaction is involved.
We may, however, take stock of the baryon number of SNs as dark matter candidates, $A\geq 10^{24}[\rho_{\rm DM}/(0.3~\rm{GeV\cdot cm^{-3}})][\varphi/(10^{-17}~\rm{cm^{-2}\cdot s^{-1}})]^{-1}$, if the the flux of SNs, $\varphi$, is lower than the constraint from IceCube's negative detection of magnetic monopoles~\citep{Icecube+2014}, with $\rho_{\mathrm DM}$ the dark matter density around the solar system.
We will hence examine SNs with baryon number, $10^{25}<A<10^{35}$, in the following sections.

We are studying the interaction of SNs travelling through the Earth or the Moon at a non-relativistic velocity, and trying to detect the signals with an acoustic array.
We perform our calculation using spontaneously magnetized SnNs~\citep{2016ChPhC..40i5102L} as an example, but the results could also be meaningful for other kinds of SNs if the cross sections are adjusted.
The physical properties such as radius, mass, number density, and cross section of SNs are introduced in Section \ref{sect:Physical}.
Section \ref{sect:AcousticDetection} is relevant to the acoustic detection of an SN, including the penetration depth, the characteristics of the sound waveform generated, and the decay and critical detectable distance of acoustic signals in water and rock media.
Discussion and conclusion are presented in Section \ref{sect:discussion} and Section \ref{sect:conclusion}.
Unless otherwise stated, we formulate in the cgs Gaussian unit system.

\section{Physical properties of strong nuggets}
\label{sect:Physical}

There are some similar properties among different kinds of SNs.
The density of SN is about 1.5 times the density of nuclear matter~\citep{Yuan+etal+2025}, $\rho_{\rm {SN}}\simeq 4\times 10^{14} \, \rm{g\cdot cm^{-3}}$, with a uniform density at pressure free.
We will use $A=10^{30}$ as a typical baryon number of the SN, and can then calculate the mass of the SN, $m_{\rm {SN}}= 1.7 \times 10^6\,\rm g\cdot(A/10^{30})$, as well as the geometric radius,
\begin{equation}\label{GeometricRadius}
    \begin{split}
	&r_0=(\dfrac{3m_{\rm {SN}}}{4\pi \rho_{\rm {SN}}})^{1/3}\\
	&= 1.6 \times 10^{-3}\, \rm{cm} \cdot (\dfrac{\it {A}}{10^{30}})^{1/3}( \dfrac{\rho_{\rm {SN}}}{10^{14} \, \rm{g\cdot cm^{-3}}})^{-1/3}.
	\end{split}
\end{equation}

Another important parameter is the charge-mass ratio of the nugget, $\delta_{\rm e}=N_{\rm e}/A$, with $N_{\rm e}$ the number of electrons in a single SN.
For strange quark matter, we can calculate that $\delta_{\rm e}\sim 10^{-6}$ to $10^{-5}$ in the MIT bag model, depending on the strange quark mass $m_{\rm s}$ (taken as $\sim 95$ MeV) and the coupling constant $\alpha_{\rm c}$~\citep{Farhi+Jaffe+1984}.
For flavour-asymmetric $ud$QM in bulk, one could also have a relatively small $\delta_{\rm e}$ to allow for Landau's neutralization.
For SnM, the ratio $\delta_{\rm e}$ could be of order of $10^{-6} - 10^{-5}$, since it  would be considered as linked bags of strangeons~\citep{Miao+2022}.
Thanks to this extremely small $\delta_{\rm e}$, SNs can be candidates for dark matter.
In this picture, most SNs in the vicinity of the solar system are virially bound to the Milky Way, moving at a typical velocity of $v=220\,{\rm km}\cdot{\rm s}^{-1}$ relative to Earth in all directions.
A small part of the SNs are generated in the supernova or compact star merger process, they may have speed up to order of $v=0.01c$~\citep{0.1c}. The number density of SN is $n_{\mathrm {SN}}=\rho_{\mathrm {DM}}/m_{\mathrm {SN}}=3.2 \times 10^{-31}\, \rm{cm}^{-3} \cdot (\it {A}/\rm{10^{30}})^{-1}$. Then the event rate of SN impact is 
\begin{equation}\label{EventRate}
    \begin{split}
	&\varphi_{\mathrm {SN}}=n_{\mathrm {SN}}v\\
	&=7.0 \times 10^{-24}\, \rm{s^{-1}cm^{-2}} \cdot (\dfrac{\it {A}}{10^{30}})^{-1} \cdot \dfrac{\it {v}}{220\,\rm{km\cdot {s^{-1}}}}.
	\end{split}
\end{equation}
Substituting the Earth's cross section $\pi R_ {\oplus}^2=1.3 \times 10^{18} \, \rm{cm}^2$, the numerical factor of the front corresponds to an event rate of about 280\,yr$^{-1}$.

The next issue which matters is to calculate the cross section of SN. First, let us look at the self-interacting cross section as dark matter.
Because of ionization, the SNs travels in the interstellar medium, with an effective interacting length order of the Debye length, $\lambda_{\rm D}=\sqrt{(k_{\rm B} T)/(4\pi n_{\rm e} e^2)}=6.9\,{\rm cm}\sqrt{T/n_{\rm e}}$.
Typically, for the interstellar medium, the temperature $T=10^4$\,K and the electron number density $n_{\rm e}=1$\,cm$^{-3}$, and one has then,
\begin{equation}\label{Debye+section}
    \begin{split}
    &\sigma_{\mathrm{SN\_self}}/m_{\mathrm {SN}}=\pi \lambda_{\rm D}^2/m_{\mathrm {SN}}\\
    &=0.9\, \rm{cm}^2 \cdot \rm{g}^{-1} \cdot (\dfrac{\it {A}}{10^{30}})^{-1},
    \end{split}
\end{equation}
which is consistent with the observational limits of self-interacting dark matter~\citep{SIDM1,SIDM2,DarkMatterCrossSection1,DarkMatterCrossSection2}.

Then, the geometric cross section of an SN reads,
\begin{equation}\label{geometric+section}
    \begin{split}
    &\sigma_{\mathrm{SN\_geo}}=\pi r_0^2\\
    &=8.0 \times 10^{-6}\, \rm{cm}^2 \cdot (\dfrac{\it {A}}{10^{30}})^{2/3} ( \dfrac{\rho_{\rm {SN}}}{10^{14} \, \rm{g\cdot cm^{-3}}})^{-2/3},
    \end{split}
\end{equation}
which is much smaller than $\sigma_{\mathrm{SN\_self}}$. It does not conflict with dark matter hypothesis.

For different kinds of SNs, because of the possible magnetic field, different types of interactions (some kinds of SN may interact with ordinary matter through nuclear force, some through electromagnetic force), and the incompletely momentum transfer from SN to medium particles, the effective cross section could be different. See Appendix~\ref{Appendix:crosssection} for detailed calculations. For convenience, we may define the cross section in general
\begin{equation}\label{sigmaSN}
    \sigma_{\rm SN}=k\cdot{(\dfrac{\it {A}}{\rm{10^{30}}})}^{2/3}\cdot 8.0\times10^{-6}\, {\rm cm}^2.
\end{equation}
Here, the factor $k$ measures the real cross sections relative to the geometric cross sections of different SNs.
Although the relations of $\sigma_{\rm SN}-A$ change in different SN models, we can still scale the cross section with $A^{2/3}$, which is proportional to the geometric section of SN,
and roughly set $k$ independent of $A$.
Certainly, the cross section $\sigma_{\rm {SN}}$ depends also on the velocity $v$, but for the sake of simplify, we treat $\sigma_{\rm {SN}}$ as a constant when $v$ changes.
This is due to the fact that the penetration depth of SN, as described in Section \ref{sect:AcousticDetection}, is greater than 10 km. For an acoustic array whose depth is a few kilometres at most, the velocity $v$ of SN does not change significantly.

The range of $k$-value is discussed as following.
On the one hand, the geometric cross section is a benchmark for the value of $k$. For an SN that interacts directly with ordinary matter via its geometric cross section, $k=1$. The incompletely momentum transfer would reduce the effective cross section, but not by orders of magnitude, so the lower limit of $k$ is taken to be $10^{-1}$. On the other hand, the cross section of the $ud$-quark nugget with magnetosphere is assumed to be the largest in all SN models, indicating the upper limit of $k$-value, 
$k=10^3$. Therefore, we take $10^{-1}<k<10^3$ in the calculations below. Table~\ref{Tab:crosssection} shows the $k$ values for different kinds of SNs.

\begin{table}
\begin{center}
\caption[]{ The $k$ Value for Different Kinds of SNs.}\label{Tab:crosssection}

 \begin{tabular}{cc}
  \hline\noalign{\smallskip}
Type of the SN  &  $k$  \\
  \hline\noalign{\smallskip}
strange quark/antiquark nugget  &  $10^1\sim10^2$  \\ 
$ud$-quark/antiquark nugget  &  $\sim10^3$  \\
axion-quark/antiquark nugget  &  $0.1\sim10^3$  \\
strangeon nugget  &  $10^1\sim10^2$ \\
  \noalign{\smallskip}\hline
\end{tabular}
\end{center}
\end{table}

\section{Acoustic detection}
\label{sect:AcousticDetection}

As SNs travel at non-relativistic velocities when they impact the Earth, the shower effect may not be significant. However, the mechanical shock wave (i.e. the acoustic wave) may be detectable. Due to the geometric radius of an SN (about $10^{-5}\sim10^{-1}\,$cm, see Eq.~\ref{GeometricRadius}) is much larger than the typical distance between molecules (about $10^{-8}\sim10^{-7}\,$cm), we can treat the medium as continuous, the hydrodynamic equations can be applied.

First, we calculate the penetration depth of an SN.
The energy deposition in unit penetration depth is $-{\mathrm d E_{\rm k}}/{\mathrm d x}=F_{\rm {drag}}=\sigma_{\rm {SN}} \cdot \rho_\mathrm {med} v^2/2$, with $E_{\rm k}=m_{\mathrm{SN}}v^2/2$.
Solving this ordinary differential equation, we can obtain
$x=4.2\times 10^{11}\, {\rm cm} \cdot k^{-1} (\dfrac{\it {A}}{\rm{10^{30}}})^{1/3} (\dfrac{\rho_{\rm med}}{1\, \rm{g\cdot cm^{-3}}})^{-1} \ln (\dfrac{220\,{\rm km\cdot s^{-1}}}{v(x)})$.
It is evident that the velocity decays exponentially with penetration depth, and the characteristic decay depth is thus
\begin{equation}\label{depth}
    \begin{split}
    x_{\rm {pene}}&=4.2\times 10^{11}\, {\rm cm} \\
    &\cdot k^{-1} (\dfrac{\it {A}}{\rm{10^{30}}})^{1/3} (\dfrac{\rho_{\rm med}}{1\, \rm{g\cdot cm^{-3}}})^{-1}.
    \end{split}
\end{equation}
As a comparison, the radius of the Earth is $6.4\times 10^8\,$cm.

When SNs impact the Earth, they will first pass through the atmosphere. To evaluate whether an SN can reach the ground, we make the approximation that the atmosphere has uniform density $\rho_{\rm med}=1.2 \times 10^{-3}\, \rm{g\cdot cm^{-3}}$ and has a thickness of $x_{\rm atm}=k_{\rm B}T/\bar m_{\rm atm}g
=8.8\,$km. Let $x_{\rm {pene}}>x_{\rm atm}$, one can derive easily that 
$A>k^3\cdot 1.6\times 10^{4}$. All the SNs with $10^{25}<A<10^{35}$ and $10^{-1}<k<10^3$ can reach the ground, we ignore the influence of the Earth's atmosphere below.

\begin{figure}[htbp]
	\centering
	\includegraphics[width=0.47\textwidth, angle=0]{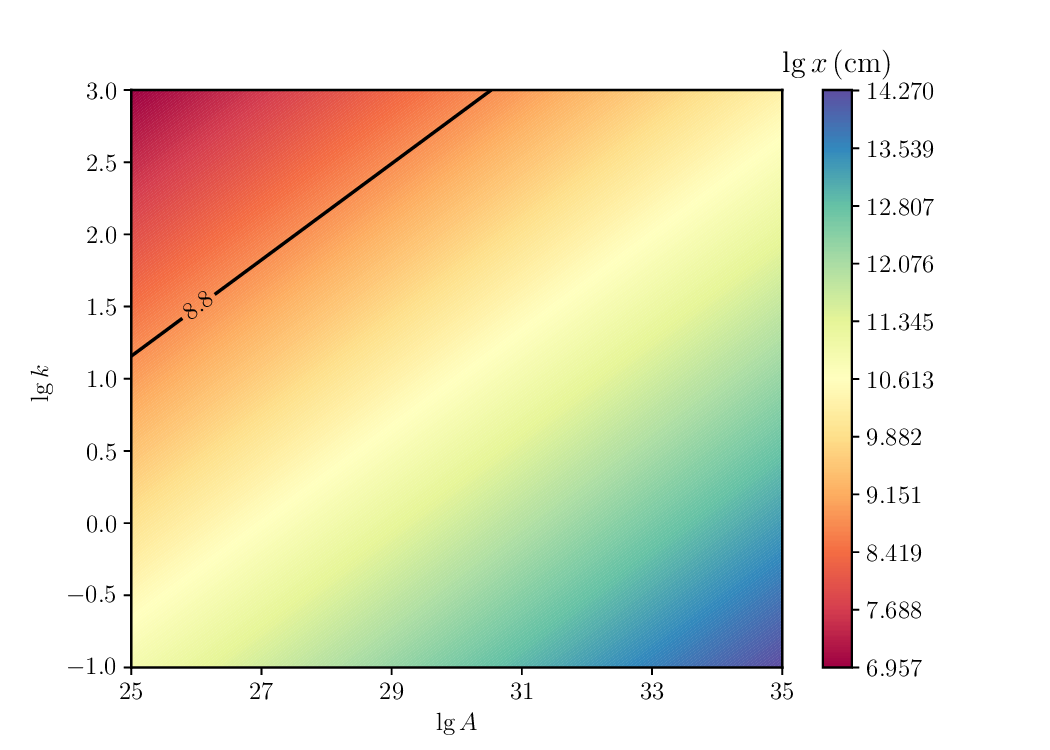}
	\caption{The characteristic penetration depth of impacting SNs. As a comparison, the radius of the Earth is about $10^{8.8}$\,cm, the solid line in the plot.}
	\label{xAk}
\end{figure}

The penetration depth based on Equation~(\ref{depth}) is shown in Figure \ref{xAk}.
SNs with small baryon number and large cross section will stay inside the Earth or the Moon after impact. However, most of SNs would easily pass through the Earth of the Moon, leaving only some energy on the trajectory, then return to the interstellar medium and re-virialise.

Second, we analyse the waveform of the SN acoustic signal.
When the SN passed, the compressed medium will generate a primary shock wave followed by a secondary wave caused by diffusion and scattering.
Using an acoustic array, we can try to detect the particular acoustic signal from SN, integrate the sound pressure signal to calculate the sound energy at different positions and times, and reconstruct the velocity and deposited energy of an SN.

Let us see the frequency of the sound wave generated by SN. Define $t_{\rm s}$ as the time interval between the pressure peaks and troughs, i.e. the duration of shock wave. \cite{Maher+2006} gives the duration from the gunshot measurement data,
\begin{equation}\label{ts}
	\begin{split}
	t_{\rm s} &=1.82\cdot\frac{2\sqrt{\sigma_{\rm SN}}}{c_{\rm s}}(\dfrac{vR}{c_{\rm s}\cdot \sqrt{\sigma_{\rm SN}}})^{1/4}\\
	&= 0.031\,{\mathrm {ms}}\cdot k^{3/8} (\dfrac{A}{10^{30}}\dfrac{R}{1\,{\mathrm {km}}}\dfrac{v}{220\,{\mathrm {km\cdot s^{-1}}}})^{1/4}\\
	&\cdot(\dfrac{c_{\rm s}}{1\,{\mathrm {km\cdot s^{-1}}}})^{-5/4},
	\end{split}
\end{equation}
where $\sqrt{\sigma_{\rm SN}}$ is the effective size of the SN ``bullet'' in the ordinary medium, $R$ is the distance from the SN trajectory, $c_{\rm s}$ is the sound speed in the medium. The frequency of shock wave is order of $f_{\rm s}=1/t_{\rm s}$.
For $10^{25}<A<10^{35}$, $10^{-1}<k<10^3$, $R=1\,$km, and $c_{\rm s}=3\,{\rm km\cdot s^{-1}}$, we have $f_{\rm s}$ from $\sim 0.54$\,kHz to $\sim 5.4$\,MHz.

We can estimate the wave frequency from another method. If we have a Gaussian wave package first, its spectrum will also be Gaussian, proportional to $\exp[-(\omega-\omega_0)^2/2\Delta^2]$, in which $\omega_0$ is the initial peak frequency of the spectrum, $\Delta$ is the spectral width (in fact, the full width of the half of the maximum is 2.355$\Delta$). In a half-cycle pulse, $\omega_0\simeq\Delta\simeq1/t_{\rm s}$. For an SN, the initial pulse frequency is about $c_{\rm s}/\sqrt{\sigma_{\rm SN}}\simeq 100\,$MHz, approaching a delta function. For a wave with a quality factor $Q$, the decay factor is $\exp(-\omega t/Q)$. If $Q$ remains constant as the amplitude decays, the peak frequency is $\omega_0-\Delta^2 t/Q$. But the peak frequency cannot be negative, so we can interpret $t$ as the timescale for the decay of a wave with frequency $\omega_0$. Therefore, for a wave detected at time $t$ after the generation of the delta-function pulse, the peak frequency should be given by the expression $\omega_0=Q/t$. At a distance of 1\,km, because $c_{\rm s}\simeq 1\,{\rm km}\cdot{\rm s^{-1}}$, the propagation time is approximately 1\,s. According to the data in~\citep{PREM1981}, $Q\simeq5.8\times10^4$ for a compression wave in the crust or in the ocean. So $\omega_0\simeq 58\,$kHz. It is close to the results from Equation~(\ref{ts}). Due to the non-uniformity and diversity of the medium, more accurate values will need to be obtained from future field acoustic measurements.

Signals from SNs with larger ${\sigma_{\rm SN}}$ could have lower frequency, but the environment noise is stronger in the low-frequency region, and the event rate of a larger ${\sigma_{\rm SN}}$ is lower.
Signals from SNs with smaller ${\sigma_{\rm SN}}$ have higher frequency, but this high frequency sound wave would decay fast in the medium.
When the distance $R$ increases, the frequency decreases.
Taking all these factors into account, we can use civilian microphones with a sampling rate of 192\,kHz to detect the SN signal whose frequency is lower than 96\,kHz (half of 192\,kHz). We could also use existing earthquake monitoring stations, but the sampling rate of existing seismographs, 1\,Hz to 1\,kHz, is relatively low~\citep{SeismicSamRate04,SeismicSamRate03,SeismicSamRate01,SeismicSamRate02}.
It is necessary to develop advanced monitors to capture the SN signals at high frequency.

\begin{figure}[htbp]
	\centering
	\includegraphics[width=0.47\textwidth, angle=0]{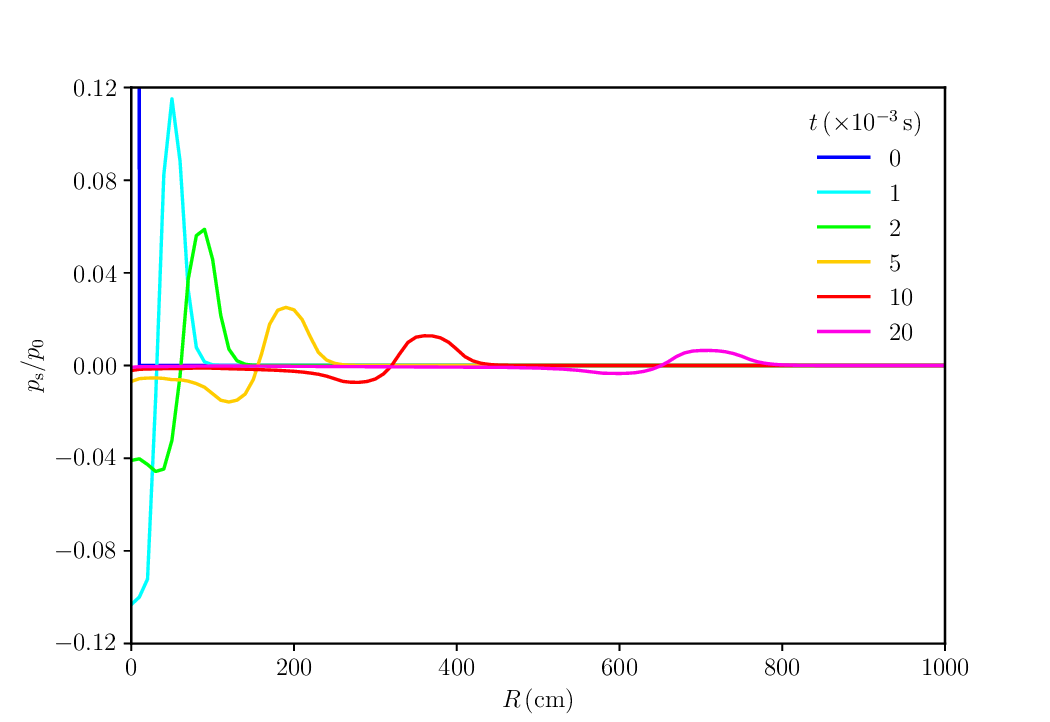}
	\caption{The hydrodynamic numerical simulation result of an instantaneous linear-shape pressure injection at $R=0$ and $t=0$. The time evolution of the waveform is plotted in the cylindrical coordinate system.
	The horizontal axis is the distance $R$ from the pressure injection line (i.e. the SN trajectory). The vertical axis is the sound pressure (i.e the deviation from hydrostatic pressure) $p_{\rm s}$, in unit of $p_0=1.0\times10^6\,{\rm erg\cdot cm^{-3}}$. As the sound wave propagates, the amplitude and the frequency decrease.}
	\label{numerical}
\end{figure}

For common medium, the sound speed $c_{\rm s}\simeq (1\sim 10) $\,km$\cdot$s$^{-1}$, the Mach number $v/c_{\rm s}$ is then much higher than 1, so the SN impact event produces an almost simultaneous pressure injection on a straight line, i.e., like a linear-shape explosion. Using a simulation code developed independently, we numerically simulate the hydrodynamics process of the pressure injection, with result shown in Figure \ref{numerical}.
The code is based on finite volume method in Godunov format~\citep{Godunov1959} of the mass, momentum and energy conservation partial differential equations, using 3-dimensional rectangular volume element. The simulation focuses on the generation, propagation, and waveform of the shock wave, so we use the inviscid conservation of momentum equations and the ideal gas equation of state to simulate the near-field behaviour of the shock wave. To reconstruct the mass density, energy density, and momentum density of the fluid from the center to the boundary of each volume element, we use piecewise linear method (PLM) with minmod gradient limiter. To calculate the conserved fluxes on each surface of each volume element, we use the HLLC solver proposed by \cite{10.1137.1025002} and \cite{HLLC}, with an adaptive improvement in the software Athena++~\citep{Athena++}. From Figure \ref{numerical} we can see the rapid reduction of amplitude of the pressure wave at a small range, mainly caused by cylindrical geometrical spreading. We can also see the frequency decreases when the sound wave propagates. The waveform shown is an obvious compression wave followed by a weaker expansion wave. These two parts make up the primary shock wave. The secondary wave is not shown because of we use an inviscid and homogeneous medium in the simulation. It is more obvious at a greater distance. We focus on the energy taken by the shock wave shown.

Third, we calculate the critical detectable distance $R_{\rm crit}$ from the impact trajectory of an SN. To find the $R_{\rm crit}$, we need to consider the influence of geometrical spreading and medium absorption of sound energy.
The relation between sound pressure $p_{\rm s}$, i.e. the deviation from hydrostatic pressure, and sound energy per unit penetration depth $\mathrm d E_{\rm s}$ in a cylindrical wavefront gives
\begin{equation}\label{dEs}
	\mathrm d E_{\rm s}=\dfrac{p_{\rm s}^2}{\rho_{\mathrm {med}}c_{\rm s}}\cdot 2\pi R \mathrm d x \cdot t_{\rm s}.
\end{equation}
The decay of the acoustic energy could be the result of medium absorption, with an exponential function at distance $R$,
\begin{equation}\label{dEsR}
    \mathrm d E_{\rm s}(R)=\mathrm d E_{\rm s0}\cdot\exp(-R/R_0),
\end{equation}
in which $\mathrm d E_{\rm s0}$ is the acoustic energy per unit penetration depth at $R\rightarrow 0$.
The value of $R_0$ 
varies with different media, temperature, and wave frequency. For example, we have $R_0=31\,$km for a 2\,kHz sound wave in $5\,^{\circ}$C ocean water, while $R_0=1.1\,$km for 20\,kHz~\citep{Sonar}.
The propagation of pressure waves in the solid crust of the Earth or the Moon is similar to that of seismic body waves. The uniformity of the crust is less than that of the ocean water, and the compression wave and the shear wave will transfer each other at crust-air or crust-mantle interfaces during propagation, so even though the crust is solid, $R_0$ will not get much larger. If the seismic wave attenuation data in Preliminary Reference Earth Model~\citep{PREM1981} is used in calculation, 
for a 2\,kHz compression wave, $R_0=30$\,km; for a 2\,kHz shear wave, $R_0=0.19$\,km.
For the purpose of estimation and averaging, we take the value of $R_0 = 1$\,km. The sound wave will travel to about several times of $R_0$, and then the noise will dominate.

\begin{figure}[htbp]
	\centering
	\includegraphics[width=0.47\textwidth, angle=0]{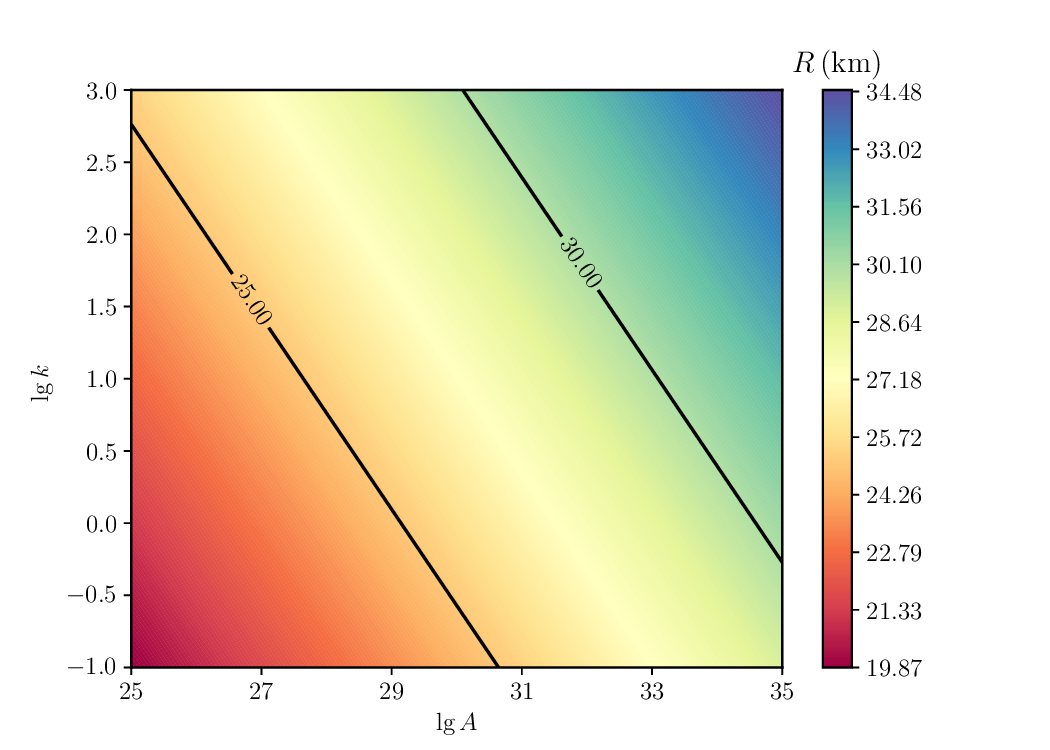}
	\caption{The critical detectable distance $R_{\rm crit}$ when $p_{\rm noise}=1\,$erg$\cdot$ cm$^{-3}$, $c_s=1\,$km$\cdot$s$^{-1}$, and $\rho_{\rm med}=1\,$g$\cdot$cm$^{-3}$. Different values of these quantities have only a small effect on the results, see Section \ref{sect:eff} for a detailed equivalent example. For different $A$ and $k$, the critical detectable distance is similar, about $25\sim30$\,km. The two reference $R_{\rm crit}$ values are shown by the solid line.}
	\label{CriticalDistancePic}
\end{figure}

We then do some more detailed calculations, taking into account the changes of $A$ and $k$. By equating the pressures of sound wave and of environment noise $p_{\rm s}=p_{\rm noise}$, one may estimate the critical detectable distance $R_{\rm crit}$. We need the drag equation due to ram pressure $-{\mathrm d E_{\rm k}}/{\mathrm d x}=\sigma_{\rm {SN}} \cdot \rho_\mathrm {med} v^2/2$, the conversion of kinetic energy into acoustic energy and decay of acoustic energy $\mathrm d E_{\rm s}=\mathrm d E_{\rm s0}\exp(-R/R_0)=(-\mathrm d E_{\rm k}) \exp(-R/R_0)$, the duration Equation~(\ref{ts}) of the sound pressure, and Equation~(\ref{dEs}) for calculating sound energy from integrating the sound pressure.
The pressure of background noise includes human and animal activity, tiny quakes, and weather. At the Salt Lake, \cite{VanDevender+2017} measured $p_{\rm noise}\simeq 1\,{\rm erg}\cdot{\rm cm}^{-3}$, about 74 dB (sound pressure level referenced to 20 \textmu Pa). We use this value to represent the background noise level around the acoustic array. The critical detectable distance $R_{\rm crit}$ satisfies,
\begin{equation}\label{CriticalDistance}
	\begin{split}
		\exp(-\dfrac{R_{\rm crit}}{1 \,\mathrm{km}}) &=\dfrac{4\pi R_{\rm crit} p_{\rm noise}^2t_{\rm s}}{\rho_{\mathrm {med}}^2 c_{\rm s} \sigma_{\rm SN} v^2} \\
		&=\exp(-16.03)\cdot (\dfrac{A}{10^{30}})^{-5/12} (\dfrac{R_{\rm crit}}{1 \,\mathrm{km}})^{5/4}\\
		&\cdot(\dfrac{p_{\rm noise}}{1\,\rm{erg\cdot cm^{-3}}})^2 (\dfrac{v}{220\,\rm{km\cdot s^{-1}}})^{-7/4}\\
		&\cdot k^{-5/8} (\dfrac{c_{\rm s}}{1\,\rm{km\cdot s^{-1}}})^{-9/4} (\dfrac{\rho_{\rm med}}{1\,\rm{g\cdot cm^{-3}}})^{-2}.
	\end{split}
\end{equation}
The calculated critical distance is shown in the Figure \ref{CriticalDistancePic}. The parameters we use are $p_{\rm noise}=1\,$erg$\cdot$ cm$^{-3}$, $c_s=1\,$km$\cdot$s$^{-1}$, and $\rho_{\rm med}=1\,$g$\cdot$cm$^{-3}$.
Different values of these parameters have only a small effect on the $R_{\rm crit}$, see Section \ref{sect:eff} for a detailed equivalent example.
We roughly have the critical detectable distance $R_{\rm crit}$ to be tens of times $R_0$.

Based on the results above, we could detect a kHz sound pulse up to 30\,km away from the trajectory of the SN using an acoustic array in the Earth's crust, the ocean, or on the Moon's surface.
Since microphones are suitable for sensing air vibrations, we need to place several microphones in a small box, and put the box in the bedrock as an acoustic detector to pick up the sound waves. In the ocean, we can use hydrophones based on the piezoelectric effect.
If we expect to detect one such SN per year, then from Equation~(\ref{EventRate}) we need a detection area of $S=(670\,{\rm km})^2\cdot (A/10^{30})$. To reconstruct the SN trajectory, we need to set up a two-dimensional acoustic array. Considering the critical detectable distance $R_{\rm crit}\simeq30$\,km, we can use a distance of 10\,km between adjacent acoustic detectors to record the propagation process of the sound wave, and use coincidence measurement on different detectors to increase the credibility. In terms of the depth, different approaches are taken depending on the medium. 
In the ocean, the depth of influence of surface waves caused by most weather, including the typhoons, is less than 100\,m. So we can place the hydrophones deeper than 100\,m. We may choose a region in the South China Sea to build the array. On the ground, we can choose an acoustically quiet location and place the microphone box at a depth of 10\,m to 100\,m to reach the bedrock. We could select the unpopulated region in Taklamakan Desert or Qinghai-Tibetan plateau. On the Moon, we could place the microphone box at a depth of several meters using robotic arm of the lunar rover.
We can process the data from the acoustic array in real time to find the possible SN shock wave signal and then discard the data without signal. The data processing algorithm can be referred to the coherent de-dispersion algorithm on radio astronomy. By using this acoustic method, we can test the hypothesis of SN-like dark matter.

\section{Discussion}
\label{sect:discussion}

\subsection{Distinction of SN signals}

An SN is very fast with very little loss of velocity. The velocity of the SN is typically $v=220\,{\rm km}\cdot{\rm s} ^{-1}$, and the penetration depth, i.e. the length of the trajectory, is comparable to the radius of the Earth. This allows us to distinguish the SN signal from other acoustic events such as weather, creature activity, cosmic dust (very fast little meteorite from outside the solar system) or quake.

The sound waves of the weather and the activities of creatures on the Earth are confined to the vicinity of the site generated. They do not have the characteristics of moving very fast along any particular long trajectory. We can eliminate these signals by reconstructing the propagation path of the sound wave using data from different acoustic detectors.

The cosmic dust could also be very fast, caused by supernova acceleration~\citep{0.1c}. But the cross section of the cosmic dust is much larger, so the penetration depth is small. One can calculate that $E_{\rm k}=m_{\rm dust}v^2/2$, $F_{\rm dust}=\pi r_{\rm dust}^2 \rho_{\mathrm {med}}v^2$, and
\begin{equation}\label{dustdepth}
    \begin{split}
	x_{\rm pene}&=110\, {\rm cm}\\
	&\cdot (\dfrac{\it {A}}{\rm{10^{30}}})^{1/3} (\dfrac{\rho_{\rm {dust}}}{3\, \rm{g\cdot cm^{-3}}})^{2/3} (\dfrac{\rho_{\rm med}}{1\, \rm{g\cdot cm^{-3}}})^{-1},
	\end{split}
\end{equation}
so the penetration depth is far smaller than that of the SN. The cosmic dust whose baryon number $A=10^{30}$ ($\sim1.7$\,tons) could become a meteor burnt in the Earth's atmosphere. On the Moon, a cosmic dust signal is more like a point-source shock wave, not a line-shape-source one. Also, because of the large cross section, the sound wave frequency of a cosmic dust is about 20\,Hz, much lower than that of an SN.

As for earthquakes and moonquakes, although they may have a long specific fault zone to generate the sound wave, but the speed of seismic rupture is roughly equivalent to the speed of sound in a cracked rock, about $(3\sim 7)\,\rm{km\cdot s^{-1}}$. It is much slower than the movement of the SN.

\subsection{Detection efficiency of the acoustic energy}\label{sect:eff}

As the SN passes through and compresses the medium, some of the kinetic energy deposited is used to generate a sound wave, and some is used to heat and ionise the medium. In addition, the sound wave will reflect on the surface of the microphone box or the hydrophone. We can define a factor $\eta$ to describe the energy transfer efficiency from kinetic energy lost by the SN to detected acoustic energy, the energy transfer relation
$\mathrm d E_{\rm s}=-\mathrm d E_{\rm k}$
is then changed to $\mathrm d E_{\rm s}=-\eta\cdot\mathrm d E_{\rm k}$, the Equation~(\ref{CriticalDistance}) becomes
\begin{equation}\label{CriticalDistance2}
	\begin{split}
		\exp(-\dfrac{R_{\rm crit}}{1 \,\mathrm{km}}) &=\dfrac{4\pi R_{\rm crit} p_{\rm noise}^2t_{\rm s}}{\eta \rho_{\mathrm {med}}^2 c_{\rm s} \sigma_{\rm SN} v^2} \\
		&=\dfrac{1}{\eta}\exp(-16.03) (\dfrac{A}{10^{30}})^{-5/12} (\dfrac{R_{\rm crit}}{1 \,\mathrm{km}})^{5/4}\\
		&\cdot(\dfrac{p_{\rm noise}}{1\,\rm{erg\cdot cm^{-3}}})^2 (\dfrac{v}{220\,\rm{km\cdot s^{-1}}})^{-7/4}\\
		&\cdot k^{-5/8} (\dfrac{c_{\rm s}}{1\,\rm{km\cdot s^{-1}}})^{-9/4} (\dfrac{\rho_{\rm med}}{1\,\rm{g\cdot cm^{-3}}})^{-2}.
	\end{split}
\end{equation}

\begin{figure}[htbp]
	\centering
	\includegraphics[width=0.47\textwidth, angle=0]{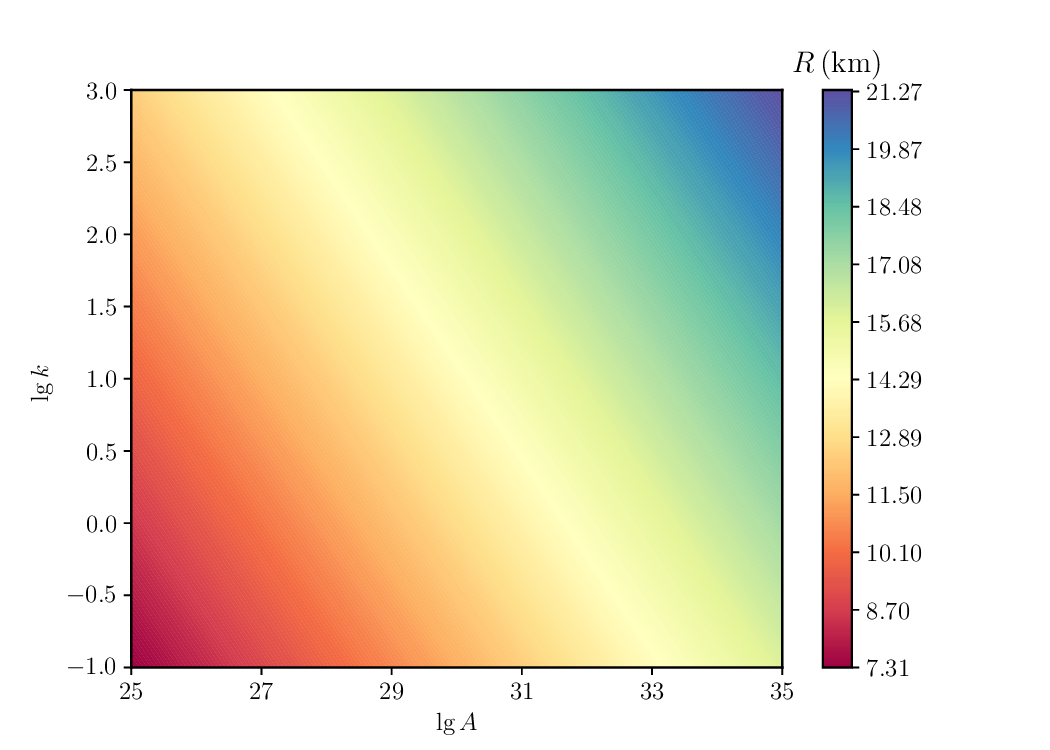}
	\caption{The critical detectable distance $R_{\rm crit}$ when the energy transfer efficiency $\eta$ from kinetic energy lost by the SN to detected acoustic energy is $10^{-6}$. It is about 1/3 to 1/2 of $R_{\rm crit}$ when $\eta=1$.}
	\label{CriticalDistancePic2}
\end{figure}

However, the influence of $\eta$ on the critical detectable distance $R_{\rm crit}$ is small. Figure \ref{CriticalDistancePic2} shows the critical detectable distance when $\eta=10^{-6}$, it is about half of the $R_{\rm crit}$ when $\eta=1$. Mathematically speaking, the change of $\eta$ is equivalent to the change of $p_{\rm noise}$, $v$, $c_{\rm s}$, or $\rho_{\rm med}$, so the different values of them will not have a significant impact on the $R_{\rm crit}$.

\subsection{The accumulation of SNs in the Earth or the Moon}

We can place an upper bound on the accumulation rate and the total accumulated mass of SNs during the Earth's 4.6\,Gyr lifetime. From Figure \ref{xAk}, SN with about $A\leq10^{30}$ could not pass through the Earth and may sink to the center of the Earth. From Equation~(\ref{EventRate}), we can calculate that the accumulation rate is less than $4.8\times10^8\,{\rm g\cdot yr^{-1}}$, the total accumulated mass is $2.2\times 10^{18}\,{\rm g}\simeq3.7\times10^{-10}M_\oplus$. The radius of the ``SN core'' in the Earth is $17$\,cm. The gravitational effect is tiny. The accumulation rate in the Moon is about 1/50 that of the Earth.

\subsection{The acoustic detection on the Moon}
\label{Moon Detection}

There is no air on the Moon, so loose lunar soil may not conduct sound well. We can place the microphone box at a depth of a few meters, using solid lunar rocks to conduct the sound wave.
Additionally, we could also learn from the placement method of existing seismographs on the Moon.

\subsection{Other uses of the acoustic array}

The event rate of the SN acoustic signal is very low, but we can use the acoustic array to detect other signals, such as meteorites and earthquakes. The acoustic array can detect acoustic vibrations when a meteorite hits the ground, or fill in the 1\,kHz to 96\,kHz band in seismic monitoring.

For meteors, there are many observatories that use radar or optical means to monitor them~\citep{Radar2006,Canada2018,BRAMON2022,BRAMON2023,Hungary2023,NASA-MEM2023}. The meteors can also generate acoustic shock waves in the atmosphere~\citep{MeteorShock}. The larger of these meteors will leave behind debris that will fall to the ground and become meteorites. The acoustic array can be built in the same area and detects meteorites in conjunction. The total mass of meteors hitting the Earth is about $2\times 10^{10}\,{\rm g}\,{\rm yr^{-1}}$~\citep{MeteorAcc2017}. A meteor heavier than $800\,{\rm g}$ may leave a meteorite~\citep{LeaveMeteorite01}. Based on the research on meteor mass-luminosity relation and the mass distribution of meteors~\citep{LumEff02andDistr,LumEff03,LumEff01}, the event rate of meteorite hitting the Earth is about $10^{5}\,{\rm yr^{-1}}$, 2 to 3 orders of magnitude higher than the event rate of SN impact. We can expect that the acoustic array can detect some signals from meteorites. 

For earthquakes, the existing seismic stations operate in the frequency band less than or equal to 1\,kHz~\citep{SeismicSamRate04,SeismicSamRate03,SeismicSamRate01,SeismicSamRate02}. The acoustic array can extend the seismic detection band up to 96\,kHz. The  distributed acoustic sensing (DAS) technology can combine acoustic and seismic detection~\citep{DAS01,DAS02,DAS03}. It can be used to acquire the Earth's motion field over a certain area, which is related to SN impact events and earthquakes. Some studies have focused on approaches to placing acoustic sensors so that they can acquire data more efficiently~\citep{SeisAcousSet01,SeisAcousSet02}.

\subsection{The budget estimation}

First we need to clarify the composition of the array. Building a single $670\,{\rm km}\times670\,{\rm km}$ microphone array is unattainable, but we can build a small array that will be used first for technology validation and preliminary detection. If we build a microphone array horizontally to $100\,{\rm km}\times100\,{\rm km}$, using a 10-kilometer square grid, we need $100$ acoustic observatory sites. On the ground, each site contains a microphone box with 5 microphones at a depth of 10\,m to 100\,m in the bedrock. In the ocean, each site contains a hydrophone at a depth of 200\,m, and a buoy to place the electronics equipment and antenna. Each site includes a wireless transmission module that encodes and gives data to the antenna, a directional antenna that transmits data, and a synchronised clock accurate to 5\,\textmu s to ensure that the signals from different microphones are synchronised. We also need a data collection system for the whole array. The maximum data rate is $192\,{\rm kHz}\times16\,{\rm bits\ per\ pressure\ sample}\times100\times2=614.4\,{\rm Mbit\cdot s}^{-1}$, it is achievable with a 1\,Gbps data broadband.

The material cost is discussed next. In order to reduce costs, we can use civilian 16-bit 192\,kHz microphones. These will pick up enough sound to detect a 74\,dB SN signal. The 192\,kHz sampling rate is sufficient to detect an SN with $A\geq10^{26}$, $k\geq10$ at a 30\,km distance. One microphone is about RMB\,50, one meter of cable is about RMB\,1, one 5\,\textmu s time synchronisation module is about RMB\,100, one wireless transmission module is about RMB\,50, one directional antenna is about RMB\,50, one hydrophone is about RMB\,200, one buoy is about RMB\,50. The data collection system and data processing computer may cost RMB\,100\,k. Based on the above description, the material cost of one site is RMB\,560 to RMB\,650 on the ground, or RMB\,650 in the ocean. So the material cost of the array is about RMB\,155\,k to RMB\,165\,k.

We then need to include in the construction and maintenance costs. We need to set up 100 sites, each requiring drilling (on the ground) or submerging (in the ocean), installation and commissioning. Then we need to build and commission the data processing center of the array. These costs are much higher than the cost of materials, which can be higher than RMB\,1 million.

On the Moon, costs will be even higher. Instruments will need to be protected from cosmic rays, the Moon landing rocket itself will cost a lot, and the long-range lunar rover is difficult to design. If we build a lunar base one day, it will be more feasible to build a lunar acoustic array.

It is a mercy that the High-energy Underwater Neutrino Telescope (HUNT)~\footnote{http://hunt.ihep.ac.cn/} has been proposed, which will use sonar to localize each underwater neutrino detector.
The HUNT will contain about 2000 vertical detector strings at 150\,m horizontal intervals, each string will include at least 4 sonar localisation modules. The array design of HUNT is $6\,{\rm km}\times6\,{\rm km}\times(1\sim2)\,{\rm km}$ at a depth of greater than 1\,km. The acoustic localisation system on the HUNT can also detect the SN signal with an event rate of about $1\,{\rm yr^{-1}}\cdot (A/10^{26})^{-1}$.
The acoustic data from HUNT would certainly constrain the baryon number $A$ of SNs in the future.

\section{Conclusion}
\label{sect:conclusion}

Both relativistic cosmic rays and non-relativistic stuff (either meteorite/cosmic dust of matter condensed by the electromagnetic force or nuggets condensed by the strong force, even the primordial black holes) will bombard the Earth frequently, but the method of detection is totally different.
For the former, secondary photoelectric and muon signals could be obtained after the air shower, and the event reconstruction could be available with a cosmic ray array.
For the latter, however, only a fireball could be produced around the incident object, and we propose in this paper to detect this signal with an acoustic array.
We expect that this novel method would play an essential role in detecting dark matter with particle masses in the range of $\sim 10^{10}$\,GeV to $\sim 10^{40}$\,GeV.

In the paper, we consider that strong nugget (SN) is a candidate of dark matter, with a baryon number $A$, $10^{25}<A<10^{35}$. Such a nugget would have a direct interaction cross section with ordinary matter.
The SNs pass through the Earth and the Moon at a non-relativistic velocity, and we proposed an acoustic method to detect these SNs.
The penetration depth of an SN could be greater than $10$\,km, even larger than Earth's radius, depending on the difference of baryon number and cross section.
The shock wave generated on the trajectory of the SN is at kilohertz frequency and has a critical detectable distance of about 30\,km. Using the data from large-area acoustic array, we can reconstruct the trajectory, velocity, and energy of the SN as it passes through.
We need $(670\,{\rm km})^2\cdot(A/10^{30})$ detection area to reach an event rate of one per year.
The array can be built on the ground, in the ocean, or even on the Moon's surface.
This experiment with acoustic array could be for the direct detection of dark matter in the macroscopic regime.
The sonar localisation system of the HUNT project is proposed as a pathfinder for SN acoustic detection.

\begin{acknowledgements}

We thank Dr. Kejia Lee for some discussions about the construction of the acoustic detecting picture, Drs. Mingjun Chen and Zhen Cao for the sonar system on the large scientific facility HUNT, Dr. Lile Wang for his guidance in programming the hydrodynamic simulation, Dr. Han Yue for some details on the generation and propagation of earthquake wave in the crust, Dr. Jinhai Zhang for moonquake monitor, and Dr. Shefeng Yan for the discussion about the acoustic detection instruments.
We acknowledge also all the members in PKU pulsar group for some very meaningful and enlightening discussions during the whole research.

\end{acknowledgements}

\appendix                  

\section{The cross section of different kinds of SN}\label{Appendix:crosssection}
We are concerned about the cross section of SN interacting with ordinary matter below. As stated in Equation~(\ref{geometric+section}), the geometric cross section of an SN is
\begin{equation*}
    \begin{split}
    \sigma_{\mathrm{SN\_geo}}&=8.0 \times 10^{-6}\, \rm{cm}^2 \\
    &\cdot (\dfrac{\it {A}}{10^{30}})^{2/3} ( \dfrac{\rho_{\rm {SN}}}{10^{14} \, \rm{g\cdot cm^{-3}}})^{-2/3}.
    \end{split}
\end{equation*}

It is worth noting that an SnN, as an example, could be spontaneously magnetized~\citep{2016ChPhC..40i5102L}, and we calculate the magnetic moment of SnN then.
An SnN interacts with ordinary matter mainly through its dipole magnetic field, which is generated by the spontaneous magnetization of electrons.
We assume that the magnetic moment of strangeons (i.e. quarks) are arranged randomly and cancel each other out, so the contribution to the total magnetic field can be neglected.
In fact, most strangeons may have 2 (spin)$\,\times\,$3 (color)$\,\times\,$3 (flavor, $u d s$)$\,=\,$18 quarks and be spin-color-flavor-symmetric, they may have no spin and no magnetic moment. Even though about $\delta_{\rm e}\simeq10^{-5}$ of all strangeons are flavor asymmetric due to the large $s$ quark mass, the large mass of the strangeon will make its spin magnetic moment very low. The total magnetic moment produced by strangeons is much lower than that produced by electrons.
SnN's magnetic moment resulting from electric current is neglected due to an extremely small magnetic Reynolds number\footnote{%
However, the dynamo action in massive strangeon stars could be effective, and strangeon magnetars could then form because of the persistent electric current~\citep{Xu2024}.
}, so we will consider only the spin magnetic moment of electrons.
As an identical fermion system, the wave function of electrons must be exchange-anti-symmetrical.
For example, one can consider the two-electron wave function without spin-orbit coupling, $\Psi(r_1,r_2,s_1,s_2)=\psi(r_1,r_2)\chi(s_1)\chi(s_2)$, in which $r_1$ and $r_2$ are the space coordinates of electrons marked 1 and 2, $s_1$ and $s_2$ are the spins of electrons. If the total spin is $S = |\Vec{s}_1+\Vec{s}_2| =1$, then $\chi(s_1)\chi(s_2)$ is exchange-symmetrical, so $\psi(r_1,r_2)$ must be exchange-anti-symmetrical. In this case, $\psi(r_1,r_2)=0$ when $r_1=r_2$, the probability that two electrons are in the same position tends to zero. If the total spin is $S=0$ instead, $\psi(r_1,r_2)$ is exchange-symmetrical and need not to be zero when $r_1=r_2$. Because of the Coulomb repulsion between electrons, the interaction energy in $S=1$ case is then lower than that in $S=0$.
As a result, spontaneous magnetisation occurs in the electron system inside the SnN, reversing the spins of many electrons, giving a large total spin $S$ and leading to a large magnetic moment $\mu$.

Let us have an order-of-magnitude calculation of the magnetic field of an SnN.
Before the spontaneous magnetization, all electrons fill the Fermi sea with two spins in the momentum space, and the electrons with spin-up ($\uparrow$) and those with spin-down ($\downarrow$) have the same Fermi momentum, $p_{\rm F}=(\dfrac{3n_{\rm e}}{4\pi})^{1/3} 2\pi \hbar$. For $\delta_{\rm e}=10^{-5}$, $n_{\rm e}\simeq \delta_{\rm e} \rho_{\rm {SnN}}/m_{\rm p}$, where $m_{\rm p}=938\,{\rm MeV}\cdot c^{-2}$, $p_{\rm F}=10\,{\rm MeV}\cdot c^{-1} \gg m_{\rm e}c$.
After spontaneous magnetization, the Fermi-surfaces of $\uparrow$-electrons and $\downarrow$-electrons are separated, with the Fermi momenta of two spins as $p_{\rm F}\pm \Delta p$.
This mis-match of the Fermi-surfaces is the result of the Coulomb interaction, so that the electromagnetic energy decreases but the kinetic energy increases, which can be expressed as, respectively,
\begin{equation}\label{dEem}
	\Delta E_{\rm em}=\xi \frac{1}{2} \dfrac{e^2}{\bar {d_{\rm e}}},~~{\rm with}~\bar {d_{\rm e}}=(\dfrac{3}{4\pi})^{1/3}\cdot \dfrac{2\pi \hbar}{p_{\rm F}},
\end{equation}
\begin{equation}\label{dEk}
	\Delta E_{\rm k}=\dfrac{p_{\rm F}\cdot 2\Delta p \cdot c}{\sqrt{(m_{\rm e}c)^2+p_{\rm F}^2}}\simeq 2\Delta pc,
\end{equation}
where $\xi$ is the correction factor for using the average distance between electrons $\bar {d_{\rm e}}$ to calculate the decrease in electromagnetic energy of electrons with parallel spin.
Since the electrostatic potential is proportional to $e^2/d_{\rm e}$, the Coulomb energy increases significantly when the distance between the electrons $d<{\bar {d_{\rm e}}}$, so we generally have $\xi>1$.
The competition between the Coulomb interaction and the kinematic excitation would cause the system to reach an equilibrium, $\Delta E_{\rm em}= \Delta E_{\rm k}$, one can then has,
\begin{equation}\label{dp}
	\dfrac{\Delta p}{p_{\rm F}}=\xi (\frac{4\pi}{3})^{1/3}\dfrac{\alpha_{\rm em}}{8\pi}= 0.064\,\xi \alpha_{\rm em},
\end{equation}
where $\alpha_{\rm em}\equiv e^2/(\hbar c)\simeq 1/137$.
As $\Delta p\ll p_{\rm F}$, there is only a small fraction of electrons whose spins are parallel when the Coulomb interaction is included for an electron gas.
Therefore, the total magnetic moment reads,
\begin{equation}\label{mu}
	\mu\simeq \dfrac{4\pi p_{\rm F}^2 \cdot 2\Delta p}{4\pi p_{\rm F}^3/3}\cdot N_{\rm e}\mu_{\rm B}=0.38\,\xi \alpha_{\rm em} \delta_{\rm e} A \mu_{\rm B},
\end{equation}
where $\mu_{\rm B}=e\hbar/(2m_{\rm e})$ is the Bohr magneton and $N_{\rm e}$ is the total number of electrons in an SnN. The magnetic field on the magnetic equator of the SnN is thus,
\begin{equation}\label{B0}
	B_0=\dfrac{\xi \mu}{r_0^3}=8.4\times 10^{11}\, \rm {G} \cdot \xi\dfrac{\delta_{\rm e}}{10^{-5}}\cdot\dfrac{\rho_{\rm {SN}}}{10^{14}\, \rm{g\cdot cm^{-3}}},
\end{equation}
which is close to the typical magnetic fields on pulsars' surface\footnote{%
This mechanism of spontaneous magnetisation is different from that of conventional ferromagnetism.
The ideal fermion gas approximation is perfect for the former, extremely relativistic electrons, but not for the latter, non-relativistic electrons in metals.
Complex interactions matter in metals.
}.

When an SN impacts the ordinary matter, such as soil, water or rock,
it makes a deposition of its kinetic energy into the surrounding medium.
The way SN loses energy is by compressing the medium and transferring momentum to the molecules of the medium. Some of the energy is converted to heat, creating an ionised environment, and some of the energy creates the shock wave as an acoustic signal.
In the aerodynamic heating picture, the temperature rises, $\Delta T=v^2/2c_{\rm p}$, where $c_{\rm p}$ is the specific isobaric heat capacity.
For solid SiO$_2$, $c_{\rm p}=7.0\times 10^6\, {\rm erg\cdot (g\cdot K)^{-1}}$. Then substitute $v=220\,\rm{km\cdot s^{-1}}$, we have $\Delta T=7.0 \times 10^7\,$K, which is high enough to ionise the atoms, causing the medium surrounding the impact trajectory to become plasma.
It is worth mentioning that SNs will not absorb ordinary matter into itself. Firstly, the ``nuclear charge number'' of an SN is $Z=\delta_{\rm e}A\simeq10^{20}\sim10^{30}$. Even when considering the nuclear charge shielding caused by the Thomas-Fermi distribution of electrons, the Coulomb potential energy barrier is too high for non-relativistic ordinary atomic nuclei to overcome because the geometric radius $r_0$ of the SN is far greater than the electron Compton wavelength $\lambda_{\rm C}$. Secondly, for nuggets with strangeness, ordinary matter must to be transformed into strange matter via weak interactions before it can be absorbed. The probability of such a transformation occurring in a single collision is extremely low.

In the case of SnN, the picture of the interaction is similar to that of the Earth's magnetosphere in the solar wind.
The medium plasma move fast relative to the SnN, then due to the strong magnetic field of the SnN, a magnetosphere forms. The radius of the magnetopause is derived by the balance between magnetic pressure and plasma ram pressure. \cite{VanDevender+2017} have already calculated this similar magnetopause radius.
If the interaction of the plasma with the magnetosphere is a completely inelastic collision, the balance gives $\rho_{\rm med} v^2/2=B(r_{\rm m})^2/(8\pi)$.
Notice that for a far dipole magnetic field, $B(r)=B_0r_0^3/r^3$, one can derive
\begin{equation}\label{rm}
    \begin{split}
	r_{\rm m}&=(\dfrac{B_0^2r_0^6}{4\pi \rho_{\rm med} v^2})^{1/6}\\
	&=0.035\, {\rm cm}\cdot (\dfrac{\it {A}}{\mathrm{10^{30}}}\xi\dfrac{\delta_{\rm e}}{\rm{10^{-5}}})^{1/3} (\dfrac{\rho_{\rm med}}{1\, \rm{g\cdot cm^{-3}}})^{-1/6}\\
	&\cdot(\dfrac{\it {v}}{\rm{220\, km\cdot s^{-1}}})^{-1/3}.
	\end{split}
\end{equation}
It implies that $r_{\rm m} \gg r_0$, and $r_{\rm m}/r_0=22\cdot (\dfrac{\xi\delta_{\rm e}}{\rm{10^{-5}}}\dfrac{\rho_{\rm {SN}}}{10^{14} \, \rm{g\cdot cm^{-3}}})^\frac{1}{3} (\dfrac{\rho_{\rm med}}{1\, \rm{g\cdot cm^{-3}}})^\frac{1}{6} (\dfrac{\it {v}}{\rm{220\, km\cdot s^{-1}}})^{-\frac{1}{3}}$ is independent of the baryon number $A$, i.e. the size of SnN.
Thanks to the large $r_{\rm m}$, acoustic detection of the SN is more feasible.

The cross section of the magnetosphere of the SnN reads,
\begin{equation}\label{sigmam}
    \begin{split}
	\sigma_{\rm m}&=\pi r_{\rm m}^2\\
	&=3.8\times10^{-3}\, {\rm cm}^2 \cdot (\dfrac{\it {A}}{\rm{10^{30}}}\xi\dfrac{\delta_{\rm e}}{\rm{10^{-5}}})^{2/3}\\
	&\cdot(\dfrac{\rho_{\rm med}}{1\, \rm{g\cdot cm^{-3}}})^{-1/3} (\dfrac{\it {v}}{\rm{220\, km\cdot s^{-1}}})^{-2/3},
	\end{split}
\end{equation}
which could be regarded as the maximum effective cross section of the SnN in the ordinary matter medium. Refer to Equation~(\ref{sigmaSN})
\begin{equation*}
    \sigma_{\rm SN}=k\cdot{(\dfrac{\it {A}}{\rm{10^{30}}})}^{2/3}\cdot 8.0\times10^{-6}\, {\rm cm}^2,
\end{equation*}
Equation~(\ref{sigmam}) is equivalent to that $k=4.8\times10^2$ for SnNs. Considering that parameters $\delta_{\rm e}$ and $\rho_{\rm med}$ may vary, and the magnetosphere may not completely block particles from passing through, $k=10^1\sim10^2$.

As with the SnN, the charge-mass ratio of a strange quark nugget is also $\delta_{\rm e}=10^{-6}\sim10^{-5}$. Its electrons can also magnetize spontaneously, so $k=10^1\sim10^2$ for strange quark nugget. The acoustic signals of SnNs and strange quark nuggets are indistinguishable.
However, in the case of the $ud$-quark nugget, flavor asymmetry leads to a higher charge-mass ratio. Using the MIT bag model, we can calculate that $\delta_{\rm e}$ can be up to $10^{-2}$ for the $ud$-quark matter, which will result in a greater dipole magnetic field and a larger radius of the magnetosphere. So $k\simeq10^3$ for the $ud$-quark nugget.
For the axion-quark nuggets, since the flavor of quarks they contain is unknown and the effects of axions are unknown, their cross sections are uncertain. We take $k=10^{-1}\sim10^3$, from incompletely momentum transfer via the geometric cross section case to the largest magnetosphere case.

\bibliographystyle{raa}
\bibliography{ms2025-0136.R1reference}

\begin{thebibliography}{70}
\providecommand\natexlab[1]{#1}
\providecommand\JournalTitle[1]{#1}

\bibitem[Aartsen {et~al.}(2014)]{Icecube+2014}
Aartsen, M.~G., Abbasi, R., Ackermann, M., {et~al.} 2014, Eur. Phys. J. C, 74,
  2938

\bibitem[Ajo-Franklin {et~al.}(2019)]{DAS02}
Ajo-Franklin, J.~B., Dou, S., Lindsey, N.~J., {et~al.} 2019, Scientific
  Reports, 9, 1328

\bibitem[Baiotti(2019)]{2019PrPNP.10903714B}
Baiotti, L. 2019, Prog. Part. Nucl. Phys., 109, 103714

\bibitem[Betzler(2022)]{BRAMON2022}
Betzler, A.~S. 2022, eMetN Meteor Journal, 7, 106,
  https://www.emeteornews.net/2022/01/20/a-comparison-of-tv-and-visual-derived-population-indexes-of-some-meteor-showers/

\bibitem[Betzler(2023)]{BRAMON2023}
Betzler, A.~S. 2023, Earth, Moon, Planets, 127, 1

\bibitem[Bodmer(1971)]{1971PhRvD...4.1601B}
Bodmer, A.~R. 1971, Phys. Rev. D, 4, 1601

\bibitem[Callahan {et~al.}(2025)]{SeisAcousSet02}
Callahan, J., Monogue, K., Villarreal, R., \& Catanach, T. 2025, Geophys. J.
  Int., 240, 1802

\bibitem[Campbell-Brown \& Jones(2006)]{Radar2006}
Campbell-Brown, M.~D., \& Jones, J. 2006, MNRAS, 367, 709

\bibitem[Carlson {et~al.}(1992)]{SIDM1}
Carlson, E.~D., Machacek, M.~E., \& Hall, L.~J. 1992, ApJ, 398, 43

\bibitem[Chandrasekhar(1931)]{Chandrasekhar1931_ApJ74-8}
Chandrasekhar, S. 1931, ApJ, 74, 81

\bibitem[Chen {et~al.}(2022)]{SeismicSamRate01}
Chen, C., Wang, Y., Guo, G.~Y., {et~al.} 2022, Chin. J. Geophys. (in Chinese),
  65, 4569

\bibitem[Chen {et~al.}(2023)]{2023RPPh...86b6201C}
Chen, H.-X., Chen, W., Liu, X., Liu, Y.-R., \& Zhu, S.-L. 2023, Rep. Prog.
  Phys., 86, 026201, arXiv:2204.02649 [hep-ph]

\bibitem[Chen {et~al.}(2016)]{2016PhR...639....1C}
Chen, H.-X., Chen, W., Liu, X., \& Zhu, S.-L. 2016, Phys. Rep., 639, 1

\bibitem[Chen {et~al.}(2024)]{2024RAA....24b5005C}
Chen, S.-C., Gao, Y., Zhou, E.-P., \& Xu, R.-X. 2024, RAA, 24, 025005

\bibitem[Cirelli {et~al.}(2024)]{2024arXiv240601705C}
Cirelli, M., Strumia, A., \& Zupan, J. 2024, Dark {Matter}, preprint
  (arXiv:2406.01705v2 [hep-ph])

\bibitem[Darzi {et~al.}(2024)]{SeismicSamRate02}
Darzi, A., Halldorsson, B., Cotton, F., \& Rahpeyma, S. 2024, Soil Dyn. and
  Earthquake Eng., 183, 108798, arXiv:2407.09338 [physics]

\bibitem[Deme {et~al.}(2023)]{Hungary2023}
Deme, L., Sarneczky, K., Igaz, A., {et~al.} 2023, WGN, Journal of the
  International Meteor Organization, 51, 166

\bibitem[Drolshagen {et~al.}(2021)]{LumEff01}
Drolshagen, E., Ott, T., Koschny, D., {et~al.} 2021, A\&A, 650, A159,
  arXiv:2011.06805 [astro-ph]

\bibitem[Drolshagen {et~al.}(2017)]{MeteorAcc2017}
Drolshagen, G., Koschny, D., Drolshagen, S., Kretschmer, J., \& Poppe, B. 2017,
  Planet. Space Sci., 143, 21

\bibitem[Dziewonski \& Anderson(1981)]{PREM1981}
Dziewonski, A.~M., \& Anderson, D.~L. 1981, Phys. Earth Planet. Inter., 25, 297

\bibitem[Farhi \& Jaffe(1984)]{Farhi+Jaffe+1984}
Farhi, E., \& Jaffe, R.~L. 1984, Phys. Rev. D, 30, 2379

\bibitem[Flambaum {et~al.}(2023)]{Flambaum+etal+2023}
Flambaum, V.~V., Samsonov, I.~B., \& Vong, G.~K. 2023, Phys. Rev. D, 107,
  123501, arXiv:2303.01697 [hep-ph]

\bibitem[Flambaum {et~al.}(2025)]{Flambaum+etal+2025}
Flambaum, V.~V., Samsonov, I.~B., \& Vong, G.~K. 2025, Phys. Rev. D, 111,
  023525

\bibitem[Fowler(1926)]{1926MNRAS..87..114F}
Fowler, R.~H. 1926, MNRAS, 87, 114

\bibitem[Godunov(1959)]{Godunov1959}
Godunov, S. 1959, Sbornik Mathematics, 47(89), 271

\bibitem[Goldman {et~al.}(2020)]{SeismicSamRate03}
Goldman, M.~R., Catchings, R.~D., Stayer, L.~M., {et~al.} 2020, Data {Release}
  for the 2016 {East} {Bay} {Seismic} {Imaging} {Investigation} of the
  {Hayward} {Fault} {Zone}, u.S. Geological Survey,
  https://doi.org/10.5066/P941WN4P

\bibitem[Guo {et~al.}(2018)]{2018RvMP...90a5004G}
Guo, F.-K., Hanhart, C., Meißner, U.-G., {et~al.} 2018, Rev. Mod. Phys., 90,
  015004

\bibitem[Harten {et~al.}(1983)]{10.1137.1025002}
Harten, A., Lax, P.~D., \& Van~Leer, B. 1983, SIAM Rev., 25, 35

\bibitem[Hewish {et~al.}(1968)]{1968Natur.217..709H}
Hewish, A., Bell, S.~J., Pilkington, J. D.~H., Scott, P.~F., \& Collins, R.~A.
  1968, Nature, 217, 709

\bibitem[Holdom {et~al.}(2018)]{2018PhRvL.120v2001H}
Holdom, B., Ren, J., \& Zhang, C. 2018, Phys. Rev. Lett., 120, 222001

\bibitem[Hughes(1974)]{LumEff02andDistr}
Hughes, D.~W. 1974, MNRAS, 166, 339

\bibitem[Karliner {et~al.}(2018)]{2018ARNPS..68...17K}
Karliner, M., Rosner, J.~L., \& Skwarnicki, T. 2018, Ann. Rev. Nucl. Part.
  Sci., 68, 17

\bibitem[Kretschmer {et~al.}(2015)]{LumEff03}
Kretschmer, J., Drolshagen, S., Koschny, D., Drolshagen, G., \& Poppe, B. 2015,
  in Proceedings of the {International} {Meteor} {Conference} (Mistelbach,
  Austria: International Meteor Organization), 209, eds., Rault, J.-L.;
  Roggemans, P.

\bibitem[Lai {et~al.}(2023)]{2023AdPhX...837433}
Lai, X.-Y., Xia, C.-J., \& Xu, R.-X. 2023, Adv. Phys.: X, 8, 2137433,
  arXiv:2210.01501 [hep-ph]

\bibitem[Lai \& Xu(2009)]{LX2009}
Lai, X.-Y., \& Xu, R.-X. 2009, MNRASL, 398, L31, arXiv:0905.2839 [astro-ph]

\bibitem[Lai \& Xu(2016)]{2016ChPhC..40i5102L}
Lai, X.-Y., \& Xu, R.-X. 2016, Chin. Phys. C, 40, 095102

\bibitem[Lai {et~al.}(2019)]{2019EPJA...55...60L}
Lai, X.-Y., Zhou, E.-P., \& Xu, R.-X. 2019, Eur. Phys. J. A, 55, 60

\bibitem[Landau(1932)]{Landau1932pZS1-285}
Landau, L.~D. 1932, Phys. Z. Sowjetunion, 1, 285

\bibitem[Madiedo {et~al.}(2013)]{LeaveMeteorite01}
Madiedo, J.~M., Trigo-Rodriguez, J.~M., Castro-Tirado, A.~J., Ortiz, J.~L., \&
  Cabrera-Caño, J. 2013, MNRAS, 436, 2818, arXiv:1309.6465 [astro-ph]

\bibitem[Maher(2006)]{Maher+2006}
Maher, R. 2006, in 2006 {IEEE} 12th {Digital} {Signal} {Processing} {Workshop}
  \&amp; 4th {IEEE} {Signal} {Processing} {Education} {Workshop} (Teton
  National Park, WY, USA: IEEE), 257

\bibitem[Miao {et~al.}(2022)]{Miao+2022}
Miao, Z.-Q., Xia, C.-J., Lai, X.-Y., {et~al.} 2022, Int. J. Mod. Phys. E, 31,
  2250037, arXiv:2008.06932 [astro-ph, physics:nucl-th]

\bibitem[Moorhead {et~al.}(2023)]{NASA-MEM2023}
Moorhead, A.~V., Milbrandt, K., \& Kingery, A. 2023, Adv. Space Res., 72, 4582,
  arXiv:2405.08685 [astro-ph]

\bibitem[Nadler {et~al.}(2025)]{DarkMatterCrossSection2}
Nadler, E.~O., An, R., Yang, D., {et~al.} 2025, ApJ, 986, 129, arXiv:2412.13065
  [astro-ph]

\bibitem[Parker {et~al.}(2014)]{DAS01}
Parker, T., Shatalin, S., \& Farhadiroushan, M. 2014, First Break, 32, 61

\bibitem[Popolo {et~al.}(2020)]{SIDM2}
Popolo, A.~D., Delliou, M.~L., \& Deliyergiyev, M. 2020, Universe, 6, 222,
  arXiv:2410.06078 [astro-ph]

\bibitem[Ren \& Zhang(2020)]{2020PhRvD.102h3003R}
Ren, J., \& Zhang, C. 2020, Phys. Rev. D, 102, 083003, arXiv:2006.09604
  [hep-ph]

\bibitem[Saad {et~al.}(2024)]{SeisAcousSet01}
Saad, O.~M., Ravasi, M., \& Alkhalifah, T. 2024, Geophysics, 89, V573,
  arXiv:2405.07660 [physics]

\bibitem[Silber {et~al.}(2018)]{MeteorShock}
Silber, E.~A., Boslough, M., Hocking, W.~K., Gritsevich, M., \& Whitaker, R.~W.
  2018, Adv. Space Res., 62, 489, arXiv:1805.07842 [astro-ph]

\bibitem[Siraj \& Loeb(2020)]{0.1c}
Siraj, A., \& Loeb, A. 2020, ApJ, 895, L42

\bibitem[Stone {et~al.}(2020)]{Athena++}
Stone, J.~M., Tomida, K., White, C.~J., \& Felker, K.~G. 2020, ApJS, 249, 4,
  arXiv:2005.06651 [astro-ph]

\bibitem[Subasinghe \& Campbell-Brown(2018)]{Canada2018}
Subasinghe, D., \& Campbell-Brown, M. 2018, AJ, 155, 88, arXiv:1801.06123
  [astro-ph]

\bibitem[Tian \& Liang(2013)]{SeismicSamRate04}
Tian, X.-B., \& Liang, X.-F. 2013, Seismic {Array} {iNegrated} {Detection} for
  a {Window} of {Indian} {Continental} {Head}, international Federation of
  Digital Seismograph Networks, https://doi.org/10.7914/SN/6A\_2013

\bibitem[Toro {et~al.}(1994)]{HLLC}
Toro, E.~F., Spruce, M., \& Speares, W. 1994, Shock Waves, 4, 25

\bibitem[Tseng \& Yeh(2024)]{DarkMatterCrossSection1}
Tseng, P.-Y., \& Yeh, Y.-M. 2024, Phenomenology of {Neutrino}-{Dark} {Matter}
  {Interaction} in {DSNB} and {AGN}, preprint (arXiv:2412.08537 [hep-ph])

\bibitem[VanDevender {et~al.}(2017)]{VanDevender+2017}
VanDevender, J.~P., VanDevender, A.~P., Sloan, T., {et~al.} 2017, Sci. Rep., 7,
  8758

\bibitem[Waite(2002)]{Sonar}
Waite, A.~D. 2002, Sonar for {Practising} {Engineers}, 3rd edn. (Hoboken: John
  Wiley \& Sons Inc.)

\bibitem[Wang {et~al.}(2024)]{2024RAA....24j5012W}
Wang, W.-Y., Zhang, C., Zhou, E.-P., {et~al.} 2024, RAA, 24, 105012,
  arXiv:2405.07152 [astro-ph]

\bibitem[Witten(1984)]{Witten+1984}
Witten, E. 1984, Phys. Rev. D, 30, 272

\bibitem[Wu {et~al.}(2024)]{Wu+2024}
Wu, X., He, W., Luo, Y., Shao, G.-Y., \& Xu, R.-X. 2024, Int. J. Mod. Phys. D,
  33, 2450020

\bibitem[Xenaki {et~al.}(2025)]{DAS03}
Xenaki, A., Gerstoft, P., Williams, E., \& Abadi, S. 2025, Distributed acoustic
  sensing for ocean applications, preprint (arXiv:2502.18344 [physics.geo-ph])

\bibitem[Xia {et~al.}(2024)]{Xia-Gao-Xu2024}
Xia, C.~J., Gao, Y., \& Xu, R.~X. 2024, Strange {Matter} and {Strange} {Stars}
  (in {Chinese}) (Beijing: Peking University Press)

\bibitem[Xu(2003)]{Xu+2003}
Xu, R.-X. 2003, ApJ, 596, L59, arXiv:astro-ph/0302165

\bibitem[Xu(2020)]{2020SCPMA..6319531X}
Xu, R.-X. 2020, Sci. China: Phys., Mech. Astron., 63, 119531, s11433,
  arXiv:2003.04506 [hep-ph]

\bibitem[Xu(2024)]{Xu2024}
Xu, R.-X. 2024, Nucl. Phys. Rev., 41, 863

\bibitem[Xu {et~al.}(2006)]{2006MNRAS.373L..85X}
Xu, R.-X., Tao, D.-J., \& Yang, Y. 2006, MNRAS, 373, L85

\bibitem[Yakovlev {et~al.}(2013)]{2013PhyU...56..289Y}
Yakovlev, D.~G., Haensel, P., Baym, G., \& Pethick, C.~J. 2013, Phys.-Usp., 56,
  289, arXiv:1210.0682 [physics]

\bibitem[Yuan {et~al.}(2025)]{Yuan+etal+2025}
Yuan, W.-L., Huang, C., Zhang, C., Zhou, E.-P., \& Xu, R.-X. 2025, Phys. Rev.
  D, 111, 063033

\bibitem[Zhang(2020)]{2020PhRvD.101d3003Z}
Zhang, C. 2020, Phys. Rev. D, 101, 043003, arXiv:1908.10355 [astro-ph]

\bibitem[Zhitnitsky(2003)]{Zhitnitsky+2003}
Zhitnitsky, A. 2003, J. Cosmol. Astropart. Phys., 2003, 010,
  arXiv:hep-ph/0202161

\bibitem[Zhitnitsky(2021)]{Zhitnitsky+2021}
Zhitnitsky, A. 2021, Mod. Phys. Lett. A, 36, 2130017, arXiv:2105.08719 [hep-ph]

\end{thebibliography}

\label{lastpage}

\end{document}